\documentclass[12pt]{article}
\usepackage{epsf}
\title{High-momentum asymtotics from the
Fock--Feynman--Schwinger path integral}
\author{Yu.A.Simonov\\ Institute of Theoretical and Experimental
Physics\\ 117218, Moscow, B.Cheremushkinskaya 25,\\ Russia}

\date{}
\def\la{\mathrel{\mathpalette\fun <}}

\def\fun#1#2{\lower3.6pt\vbox{\baselineskip0pt\lineskip.9pt
\ialign{$\mathsurround=0pt#1\hfil
##\hfil$\crcr#2\crcr\sim\crcr}}}
\newcommand{\be}{\begin{equation}}
 \newcommand{\ee}{\end{equation}}

\begin{document}
\maketitle

\begin{abstract}

The asymptotics of n-point Green's function at large external momenta
is obtained in  the exponentiated form using the
Fock--Feynman--Schwinger  representation for propagators in the
external field. The method is applied to  gauge
theories such as QCD and QED, and the Sudakov form--factor is
calculated as an example in QED and meson form-factor in QCD.
Nonperturbative contributions can be conveniently included, as it is
demonstrated in the example of the confinement correction to the
form-factor.
\end{abstract}

\section{Introduction}

The high--momentum asymptotics of QCD and QED amplitudes is under
active study for several decades [1,2]. The basic method used is the
summing up of the series of dominant Feynman amplitudes keeping the
leading asymptotic term in each of them. These terms appear to be
of the type $\alpha^n_s ln^r Q^2$ and in case $r=2n$ one has the
so-called double logarithmic asymptotics, which e.g. defines
behaviour of the QED form--factor at large $Q^2$ [3]. An extension of
the double--logarithmic asymptotics  to other processes has been made
both in QED [4] and QCD [2,5].

One of attractive features of these results is the exponentiation of
leading logarithmic terms coming from each Feynman amplitude due to
summation over $n$, which might be a hint, that there can exist an
alternative, more direct method of deriving  this result.
A step in this direction is made in the present paper. We start with
a most general n-point Green's  function, depending on $n$ external
momenta $p_k, k=1,...n$. Two kinds  of theories are considered: QED
and QCD, but the generalization to the EW case is straightforward.
In this first study we confine ourselves to the simplest setting of
the problem:
 all combinations of
momenta are large compared to masses, external $M_i$  and internal
$m_i$,
\be
p^2_i\gg M_i^2, ~~m_k^2,~~i,~~k=1,...n
\ee
Moreover the following condition is assumed to hold
\be
|p_ip_k|\gg p^2_i,p^2_k; i,k=1,...n
\ee
Our main object of the study will be the one--fermion--loop
amplitudes which assumes the limit of large $N_c$ for QCD. However
one may expect that the inclusion of additional fermion loops gives
only small correction to the leading asymptotics, so that the one--
fermion--loop amplitudes are dominant in the asymtotics (1) and (2)
both for QCD and QED.

For QCD there is a special problem of nonperturbative (NP)
contributions, which can be handled as in [6], i.e. dividing the total
vector field $A_\mu=B_\mu+a_\mu$, where $B_\mu$ is the NP background
and $a_\mu$ is the perturbative field.
However with the conditions (1) and (2) the  standard
expectation is
that the NP contribution dies out as a power, e.g.
$\sigma/p^2_i,\frac{\sigma}{|p_ip_k|}$, where $\sigma $ is the string
tension (or any equivalent combination like gluonic condensate) made
of the NP field $B_\mu$.  We shall argue in the last section of the
paper that actually NP effects  can be also exponentiated
 and may  strongly modify the perturbative
result. The plan of the paper is as follows. In section 2 we
introduce FFSR for the $n$--point Green's function . In section 3 the
approximate path integration is done in a way  pertinent to the high
momentum kinematics (1),(2) and double logarithmic terms in the
asymptotics are calculated. In section 4 the Sudakov vertex
asymptotics is identified and formfactors are discussed. In section 5
the influence of confinement is considered and as a result
the confinement correction is calculated in addition to the  DLA
 term.  Discussion of other
 applications is given in
 the conclusions, and in the appendix the spin--dependent terms
$(\sigma F)$ in the fermion propagators are estimated and shown to be
subleading.  Note, that the situation with NP fields changes, if one
 abandons condition (1), since the Green's function under
consideration has bound state poles at $p^2_i=M^2_i$, and those are
mostly due to the NP forces between quark and antiquark.  As it is,
however, both conditions ensure that to the leading approximation one
can consider only perturbative (gluon or photon) exchanges between
fermion lines and the main point is how to sum up those in the most
effective way.

To this end we shall exploit the Fock--Feynman--Schwinger
representation (FFSR)  for the fermion propagator in the external
field [7,8], which was used previously for similar purposes in
[9].
The path--integral method have been used for the calculation of the
Green's function asymptotics in QED for a long time, see e.g. [10]
and refs. therein. More recently the so-called world line method was
introduced in [11]. Below we follow the formalism of [9], which
allows for a more economical treatment of spin.

 \section{Integral
representation for the fermion propagator}

For the spin--1/2 particle propagator in the external vector field
$A_\mu$ one has the FFSR
(we start with the Euclidean metric for convenience)
\be
S(x,y)=(m-\hat D)\int^\infty_0 ds(Dz)_{xy}e^{-K} \Phi_\Sigma(x,y)
\ee
where notations used are
\be
K=m^2s+\frac{1}{4}\int^s_0 d\tau(\frac{dz_\mu}{d\tau})^2;~~
D_\mu=\partial_\mu-igA_\mu,
\ee
\be
(Dz)_{xy}=lim_{N\to \infty}\prod^N_{n=1}\frac{d^4\xi(n)}
{(4\pi\varepsilon)^2}\frac{d^4q}{(2\pi)^4} e^{iq(\sum\xi(n)-(x-y))},
N\varepsilon=s
\ee
\be
\Phi_\sigma(x,y)=P_AP_F exp ig \int^x_y A_\mu dz_\mu exp g\int^s_0
d\tau\sigma_{\mu\nu}F_{\mu\nu},
\ee
and
$$
\sigma_{\mu\nu}=\frac{1}{4i}(\gamma_\mu\gamma_\nu-\gamma_\nu\gamma_\mu),
~~\xi(n)=z(n)-z(n-1).
$$
Consider now the n-point Green's function with external momenta at
$n$ vertices equal to $p_i$ (see Fig.1)
\be
G(p_1,...p_n)=<J_1(p_i)...J_n(p_n)>, J_i(x)=\psi(x)\Gamma_i\psi(x)
\ee
Insertion of (3) into (7) for the one--fermion loop yields
\be
G(p_1,... p_n)=<tr\prod^n_{i=1}\Gamma_i(m_i-\hat D_i)\int^\infty_0
ds_i(Dz^{(i)})_{x^{(i)},x^{i-1}}e^{-K_i}\Phi^{(i)}_\sigma
e^{ip^{(i)} x^{(i)}}dx^{(i)}>_A
\ee
 We shall disregard in what
follows the factors $\Gamma_i(m_i-\hat D_i)$ since we shall be
interested only in the exponentiated contributions; doing the
$dx^{(i)}$ integrals, one obtains
 $$ G\to \bar
G_n\delta(\sum^n_{i=1}p_i)(2\pi)^4,
$$
 where
 \be
  \bar
G_n=\int\frac{d^4q}{(2\pi)^4}\prod^n_{i=1}
ds_i\prod^N_{k=1}\frac{d\xi^{(i)}(k)}{(4\pi\varepsilon)^2}
e^{iq^{(i)}\sum_k\xi^{(i)}(k)}e^{-K_i}<
\Gamma_i(m_i-\hat D_i)W_\sigma>
\ee
and
$$ <W_\sigma>=<\prod^n_{i=1}\Phi_\sigma^{(i)}>_A,
$$
The integral $d^4q$ denotes the integral over one of $q_i$, all
others being expressed through it and all $p_i$.

We note that $<W_\sigma>$ is a gauge invariant quantity summing
all the perturbative exchanges  inside the fixed Wilson contour,
defined by the set $\{\xi^{(i)}(k)\}$. In addition to the usual
Wilson (charge) vertices, there are also magnetic moment vertices
$\sigma F$, hence the notation $<W_\sigma>$.

We concentrate now on the contribution of the $A_\mu$ field in (6),
referring the reader for the discussion of the $\sigma F$ term to
the Appendix.

The  crucial step for what follows is the use of the cluster
expansion theorem [12], which yields for $<W_\sigma>\to <W>$
\be
<W>\equiv exp
\{\sum^\infty_{r=1}\frac{(ig)^r}{r!}
\sum_{k_i}\xi_{\mu_1}(k_1)\xi_{\mu_2} (k_2) ... \xi_{\mu r} (k_r)\ll
A_{\mu_1} (z_{k_1})...
A_{\mu_r} (z_{k_r})\gg\}
\ee

Here double brackets denote cumulants [12], the lowest order
contribution (in the exponent) is expressed through photon (gluon)
propagator, which in the Feynman gauge is (the gauge is irrelevant
since $<W>$ is gauge invariant)
\be
<A_{\mu}(z)A_\nu(z')>= \frac{\delta_{\mu\nu}C_2(f)\hat 1}{4\pi^2
(z-z')^2}
\ee
Here $C_2(f)$ is the quadratic Casimir operator for fundamental
representation, $\hat 1$ is the unit color matrix, for QED one should
replace $C_2\hat 1\to 1$.

In what follows we confine ourselves to the lowest contribution (11)
in (10) and show that it yields the double logarithmic asymptotics,
leaving next order terms for the future.

First of all one can persuade oneself that the approximation (11)
yields in (9) all diagrams with exchanges of photon/gluon lines
between fermion lines,  all orderings of lines included. For QCD this
means the following: all orderings, i.e. all intersection of gluon
lines in space-time are included, except that color ordering of
operators $t^a$ is kept fixed. Since commutator of   any two
generators $t^a$ is subleading at large $N_c$, it means that (11)
sums up all exchanges including intersections of gluonic lines in the
leading $N_c$ approximation ($cf$ the discussion of this point in
[13],
note that e.g. the nonplanar diagram $0(g^4)$ is given by the
quartic cumulant in (10) and the latter is suppressed by the factor
$1/N_c$).

\section{The Gaussian integration}

Our next point is the integration over $d\xi(k)$ in (9) which is
Gaussian in the main term $K_i$, defining the measure of integration;
therefore we shall do it expanding the  exponent in (9) around the
stable fixed point $\bar \xi$, which is obtained by differentiating
the exponent in (9) with respect to $\xi^{(i)}(k)$. One has
$$
\bar
\xi^{(i)}(k)=2\varepsilon_i\{iq^{(i)}-
\frac{g^2C_2(f)}{4\pi^2}\sum_{j,k'} \frac{\bar \xi^{(j)} (k')}{(\bar
z^{(i)}(k)-\bar z^j(k'))^2}+
$$
\be
+\frac{2g^2C_2(f)}{4\pi^2}\sum_{j,k'} \sum_{m\geq k}\frac{(
\bar \xi^{(i)} (m)\bar \xi^{(j)}(n'))(\bar z^{(i)}(m)-\bar
z^{(j)}(n'))}{(\bar z^{(i)}(m)-\bar z^{(j)}(n'))^4}\}+ 0(g^4)
\ee
Here e.g. $\bar z^{(i)}(k)=\sum^i_{j=1}\sum^k_{\nu=1}\bar
\xi^{(j)}(\nu)$, where we have chosen as the origin the coordinate
$x^{(1)}$ of the first vertex, and all other coordinates are
calculated using the connection
$x^{(i)}-x^{(i-1)}=\sum^N_{k=1}\xi^{(i)}(k)$ with the cyclic
condition $x^{(n+1)}=x^{(1)}$.

One can solve the system of equations (12) iteratively  expanding in
powers of $g^2$, the first two terms are given in (12),
where one should replace $\bar \xi^{(i)}$
inside the curly brackets by $2i\varepsilon_i q^{(i)}$. If one
represents the exponential appearing in (9) after insertion of (10)
as $exp (-f(\xi, q))$, then one can write
\be
f(\xi,q) =
\sum_{i,k}\frac{(\xi^{(i)}(k))^2}{4\varepsilon_i}-i\sum_{i,k} q^{(i)}
\xi^{(i)} (k)-\frac{g^2C_2(f)}{8\pi^2} \sum_{i,j,kk'}\frac{\xi^i(k)
\xi^j(k')}{(z^i(k)-z^j(k'))^2}+0(g^4)
\ee

The Gaussian integration in (9) finally yields
\be
\bar G_n\sim \int\frac{d^4q}{(2\pi)^4} \prod^n_{i=1} ds_i e^{-f(\bar
\xi,q)-\frac{1}{2} tr ln \varphi}
\ee
where the matrix $\varphi$ is
\be
\varphi^{ij}_{kn} = \frac{1}{2} \frac{\partial^2}{\partial
\xi^{(i)}(k) \partial\xi^{(j)} (n)}f(\xi, q)\biggl | _{\xi=\bar \xi}
\ee

The most important  for what follows is the term $f(\bar
\xi,q)$ which can be written as (at this point we
reestablish Minkowskian metric)
\be
f(\bar \xi, q) =\sum^n_{i=1}s_i(q^{(i)})^2+\frac{g^2C_2 (f)}{8\pi^2}
\sum_{ij}\int^{s_i}_0
\int^{s_j}_0\frac{d\tau_id\tau_j(q^{(i)}
q^{(j)})}{(\tau_iq^{(i)}-\tau_jq^{(j)}
-\Delta_{ij})^2}
\ee
where we have defined $\tau_i=k\varepsilon_i$, and
\be
\Delta_{ij}=\sum^{j-1}_{k=i}s_kq^{(k)}, i<j
\ee

The integral in the last term on the r.h.s. of (16) can be written as
 $s_is_j(q^{(i)}q^{(j)})I(s,q)$, where
 \be
 I_{ij}(s,q)=\int^1_0\int^1_0\frac{d\alpha
 d\beta}{(\alpha s_iq^{(i)}-\beta
 s_jq^{(j)}-\Delta_{ij})^2}
 \ee
 The diagonal terms, $I_{ij},$ with $i=j$ do not contribute to the
 asymptotics and contain only  selfenergy divergencies, which are
 of no interest to us in what follows. Therefore we shall consider
 nondiagonal terms with $i\neq j$.

 Let us first study the term with $i=j-1$ ('the dressed
 vertex contribution"), and $\Delta_{i,i+1}=s_iq^{(i)}$, see Fig. 2.
 Then (18) is reduced to the form which will be studied
 below
 \be
 I_i\equiv I_{i,i+1}(s,q)=\int^1_0\int^1_0
 \frac{d\alpha
 d\beta}{(\alpha s_iq^{(i)}+\beta
 s_jq^{(j)})^2}=
 \int^1_0\int^1_0
 \frac{d\alpha
 d\beta}{(a^2\alpha^2+\beta^2b^2+2\alpha\beta(ab)}
 \ee
 with $a=s_iq^{(i)}, b=s_jq^{(j)}, j=i+1$.
 As it stands the integral (19) diverges at small $\alpha,\beta$ (or
 at small $\tau_i,\tau_j$ in (16)). The origin of this divergence
 becomes physically clear, when one expresses the distance $z^{(i)}$
 from the vertex position,
 (we go over to the Minkowskian space--time)
 \be
 z^{(i)}=2q^{(i)}\tau_i, z^{(j)}=2q^{(j)}\tau_j.
 \ee
 The quasiclassical motion (20) cannot be true for small $\tau_i$
 when quantum fluctuations wash out the straight--line trajectories,
 and the lower limit $\tau_{min}$ can be obtained from
  the quantum uncertainty principle
 \be
\Delta z\Delta q\sim (z^{(i)}-z^{(j)})(q^{(i)}-q^{(j)})\sim 1
\ee
We shall be interested in the kinematical region, where
\be
|(q^{(i)}q^{(j)}|\gg(q^{(i)})^2, (q^{(j)})^2,
\ee
 and $\tau_{min}$ then
is found from (21) to be
\be
\tau_{min}\sim \frac{1}{2|q^{(i)}q^{(j)}|}
\ee
Using (23) one can easily calculate the integral (19) since the term
$2\alpha\beta(ab)$ in the denominator of the integrand in (19) always
dominates. The result is
\be
I_i=\frac{1}{2s_is_{i+1}(q^{(i)}q^{(i+1)})}ln(2(qq')s_i)ln(2(qq')
s_{i+1})
\ee
The integration of the general term $I_{ij}$ with $j\neq i-1,i+1$
can be done using  the expressions for the Spence functions.
 However in the
general case the lower limit $\tau_{min}$ is inessential and the
double logarithmic situation does not occur unless there is
a large ratio, $|\frac{(q_iq_k)}{(q_lq_m)}|\gg 1$.
 We leave the detailed
study of this point to the future.

\section{The 3-- point Green's functions}

We start with the open triangle, Fig.2 corresponding to the Sudakov
vertex function asymptotics,
\be \bar G_3 = (-i\hat q +m)^{-1}
\Gamma(q,q')(-i\hat q'+m)^{-1}
 \ee
  In this case there is no
integration over $dq$ in (9) and only one integral $I_{12}$ (18),(24)
is present (we disregard as before the selfinteracting pieces
$I_{ii}$, which do not contribute to the asymptotics).

Insertion of the integral (24) into (16) and integration over
$ds_1ds_2$ in (14) which yields in (24) (with logarithmic accuracy)
replacement $s_i\to \frac{1}{q_i^2}$, finally leads one to the answer
for QED $(C_2\equiv 1)$
\be
\Gamma(q,q')\sim exp (-\frac{\alpha}{2\pi}ln \frac{2|qq'|}{q^2} ln
\frac{2|qq'|}{(q')^{2}})
\ee
which coincides with the known Sudakov asymptotics [3].

We turn now to the case of QCD, where the basic  triangle diagram is
closed due to color gauge invariance and try to find out whether the
kinematical region (22) plays important role in the integral over
$dq$ in (14).

In the general case, when all $q_i$ are unconstrained and expressed
through three external momenta $p_1,p_2,p_3$ and one integration
variable, the region (22), yielding double logarithmic asymptotics
(DLA) (24), is suppressed due to large values of $f(\bar \xi, q)$ in
the exponent. As a result the integral over $dq$ does not lead to the
DLA form for $\bar G_3$.

The situation changes however, if one considers instead of   $\bar
G_3$ the formfactor, i.e. when the pole terms are factored out from
the vertices 2 and 3 see Fig. 3, and vertex functions appear there.
To simplify matter, one can consider for the formfactor the same
representation (14), where under the integral one introduces vertex
functions $\psi_i(k_i),i=1,3$, where
$$
k_1=
(q^{(1)}+q^{(3)})-p^{(1)}\frac{(q^{(1)}+q^{(3)})p^{(1)}}{(p^{(1)})^2}
$$
\be
k_3=
q^{(2)}+q^{(3)}-p^{(3)}\frac{(q^{(2)}+q^{(3)})p^{(3)}}{(p^{(2)}_3) }
 \ee

The definition (29) yields in the c.m.system of particle 1 or 3 the
 familiar relative momentum of two emitted fermions. The presence of
 $\psi_i$ imposes restriction on momenta $q_i$, namely
 \be
 k^2_1, k^2_3 \la \kappa^2
 \ee
 where $\kappa^2$ is some hadronic scale.

  We can define in the Breit system momenta as follows $p^{(1)},
  p^{(2)}, p^{(3)}= (p_0,-\frac{\vec Q}{2}),(0,\vec Q), (p_0,
  \frac{\vec Q}{2})$ where $\vec Q^2\gg \kappa^2$ and
  $p_0^2=M^2+\frac{\vec Q^2}{4}$.

   One can easily see, that (30) constrains the region of integration
   over $dq\equiv d^4q_1$ to the region $|\vec q |\sim
   \kappa, |q_0-p_0|\sim \kappa$, and conditions (22) are
   satisfied.

   Hence in this case one can reproduce the Sudakov
   asymptotics (26), where
   \be
   |qq'|\to |q^{(1)}q^{(2)}|\approx \frac{\vec Q^2}{4},
   q^2\sim q^{\prime 2}\sim \kappa^2,~~
   \alpha\to \alpha_sC_2
   \ee

   \section{Modifications due to confinement}

   As was discussed in the Introduction,  in the kinematical region
   (1),(2)  one usually assumes that all distances in the $n$--point
   function are small and therefore nonperturbative interaction is
   ineffective. Here we show that this is in general not true, and in
   particular for the DLA  processes large distances appear
   naturally. Indeed, the distances between two straight--line
   trajectories (20), starting from some vertex,
    $l\equiv |z-z'|\sim
   2 |q\tau -q'\tau'|$ may  become large, $
   l=0(\sqrt{\frac{|qq'|}{q^2q^{\prime 2}}})$, at the end points,
   $\tau=s\sim\frac{1}{q^2},$ $\tau'=s'\sim \frac{1}{q^{\prime 2}}$.
   The  nonperturbative background $B_{\mu}$ produces the area--law
   term for $<W_\sigma>$ in (9), $<W_\sigma> \sim exp (-\sigma
   S_{\min})$, where $S_{\min}$ is the
   area of the minimal surface bounded by the
   line of trajectories, $\{ \xi^{(i)}(k)\}$, and the natural measure
   for influence of confinement is the value of $\sigma S_{min}$ for
   trajectories (20). For the triangle formed by the latter
   the area is  $
   S_{min}\sim \sqrt{(zz')^2}\sim |qq'|\tau\tau'$. It is clear that
   confinement starts to play role at the value
   $\tau\sim\tau'\sim\tau_0\equiv (\sigma|qq'|)^{-1/2}$ and at later
   proper time trajectories are no more straight lines. To estimate
   corrections to the trajectories one can add the term $\sigma
   S_{min}$ to the effective action (13). Differentiation of
   $\sigma S_{min}$ with respect to $\xi_\mu^{(i)}(k)$ yields an
   additional term on the r.h.s. of (12) $\Delta \xi \sim \sigma
   r^{\perp}$, where $r^{\perp}$ is the component of
   $r_\mu=2i(q_\mu\tau-\bar q_\mu\bar \tau)$ perpendicular to
   $\xi^{(i)}_\mu (k)$. Since $\tau_i\leq s_i,~~ s_i\sim
   \frac{1}{q^2_i}$, one obtains correction $O(\frac{\sigma}{q^2})$
   to the leading term $q^{(i)}$ in the curly brackets of (12).
   Hence the condition of stability of Sudakov asymptotics with
   respect to the nonperturbative (confining) effects is
   $q^2_i\gg\sigma$.

   However there is nevertheless an additional contribution due to
   $\sigma S_{min}$ in the exponent, which obtains when one inserts
   in $S_{min}$ straight--line trajectories (20). With the
   conditions (22) one gets
   \be
   <\exp(-\sigma S_{min}>_{eik} \cong exp
   (-\frac{2\sigma|q_iq_j|}{|q^2_iq^2_j|})
   \ee
   One can notice that the term in the exponent is not necessarily
   small and may substantially modify the double--logarithmic
   asymptotics.

   \section{Conclusion}

   The formalism  described above is a direct development of the
   relativistic path integral, using the proper time, which we call
   Fock--Feynman--Schwinger Representation (FFSR) [7,8]. As in other
   versions of  the path integral method [10,11] it is convenient to
   yield results already in the exponentiated form. We kept here the
   bispinor variables in the original Feynman form (the term
   $<W_\sigma>$ in (9)) which in principle allows to calculate
   spin--dependent effects also in an exponentiated form, however it
   was shown in the appendix that those effects are subdominant. One
   of the important advantages of the present formalism is that it
   allows to take into account NP effects also in the exponent, as
   it was demonstrated in section 5.

   There are possible many lines of development, including
   application to Drell--Yan and heavy--quark production, as well as
   finding connections with the factorization theorem, as it was done
   in a similar formalism in [14].

   The author is grateful to A.B.Kaidalov for useful suggestions and
   to L.N.Lipatov for discussion of the results.

\setcounter{equation}{0} \def\theequation{A.\arabic{equation}}

   \section{Appendix. Contribution of the magnetic moment term
   $\sigma_{\mu\nu}F_{\mu\nu}$.}

   We prove below that $(\sigma F)$ terms do not contribute to the
   DLA, but yield the subleading terms.

   The terms to be estimated are
   \be
   <exp (g\int^s_0d\tau \sigma_{\rho\mu}F_{\rho\mu}(z(\tau))\Gamma_j
   exp (g\int^{s'}_0d\tau'
   \sigma_{\sigma\nu}F_{\sigma\nu}(z'(\tau')))>
   \ee
   To illustrate the procedure it is enough to consider the $0(g^2)$
   terms, namely
   \be
   \sigma_{\rho\mu}\gamma_j\sigma_{\sigma\nu} \int^s_0d\tau
   \int^{s'}_0d\tau' D_{\rho\mu, \sigma\nu}
   \ee
   where $D_{\rho\mu,\sigma\nu}\equiv <F_{\rho\mu}(z)
   F_{\sigma\nu}(z')>$ can be written perturbatively as $(u\equiv
   z-z')$
   \be
   D_{\rho\mu,\sigma\nu}=\frac{1}{2\pi^2} [\frac{\partial}{\partial
   u_\rho}(u_\sigma\delta_{\mu\nu}- u_\nu\delta_{\mu\sigma})+
   \rho\sigma\leftrightarrow \mu\nu]\frac{1}{u^4}
   \ee
   Commuting $\gamma$--matrices (A.2)  can be rewritten as
   \be
   \frac{2}{\pi^2}(4\gamma_\mu\delta_{\nu j}-\gamma_j\delta_{\mu\nu})
   \int^s_0d\tau\int^{s'}_0d\tau'\frac{u_\mu u_\nu}{u^6}
   \ee
   In the lowest saddle--point approximation for $\xi$, Eq. (12), one
   can replace according to (20)
   \be
   u_\mu=2 i(q_\mu\tau+q'_\mu\tau')
   \ee
   At the same time the lower limit of integration over $\tau,\tau'$
   in (A.4) should be replaced by $\tau_{min}=\frac{1}{2|qq'|}$, and
   the integral can be estimated as
   \be
   \frac{1}{2|qq'|}[\frac{q_\mu q_\nu}{2}I_1+
   (q_\mu q'_\nu+q'_\mu q_\nu)I_2+
   \frac{q'_\mu q'_\nu}{2}I_3]
   \ee
   where $I_n, n=1,2,3$ grow at most logarithmically at large $|qq'|$
    and all terms of the order
   $0(\frac{q^2}{|qq'|}, \frac{q'^2}{|qq'|})$ have been neglected.
   Thus the terms $\sigma F$ do not contribute to DLA in the order
   $0(g^2)$. One can follow the same procedure to higher orders and
   persuade oneself that leading powers of logarithms, $\alpha
   ln^2|qq'|$, do not appear.

\newpage
\begin{figure}[thb]
\epsffile{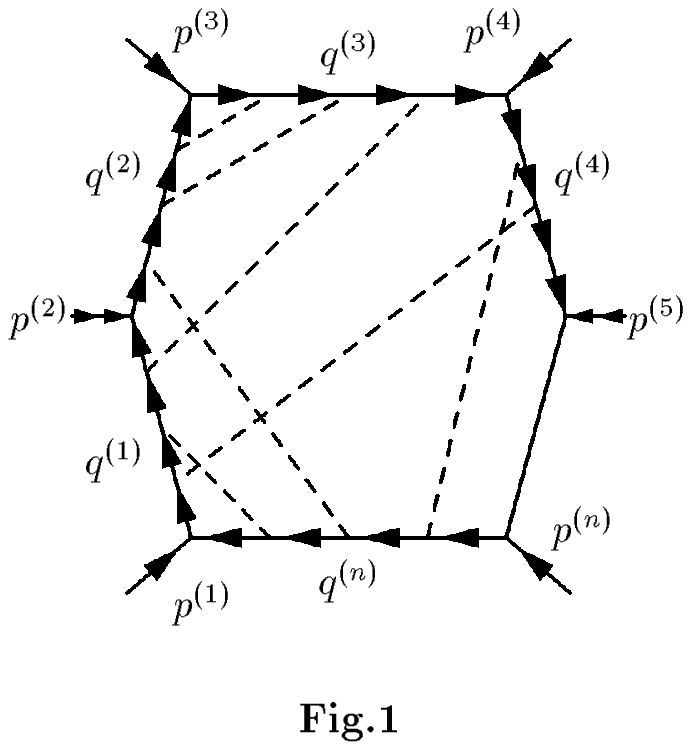}
\end{figure}
\newpage
\begin{figure}[thb]
\epsffile{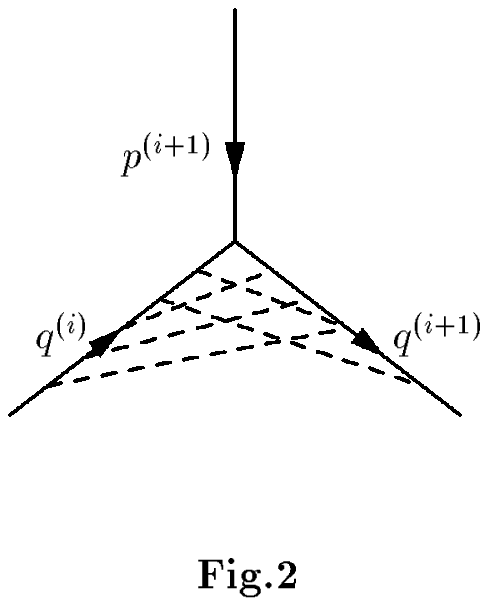}
\end{figure}
\newpage
\begin{figure}[thb]
\epsffile{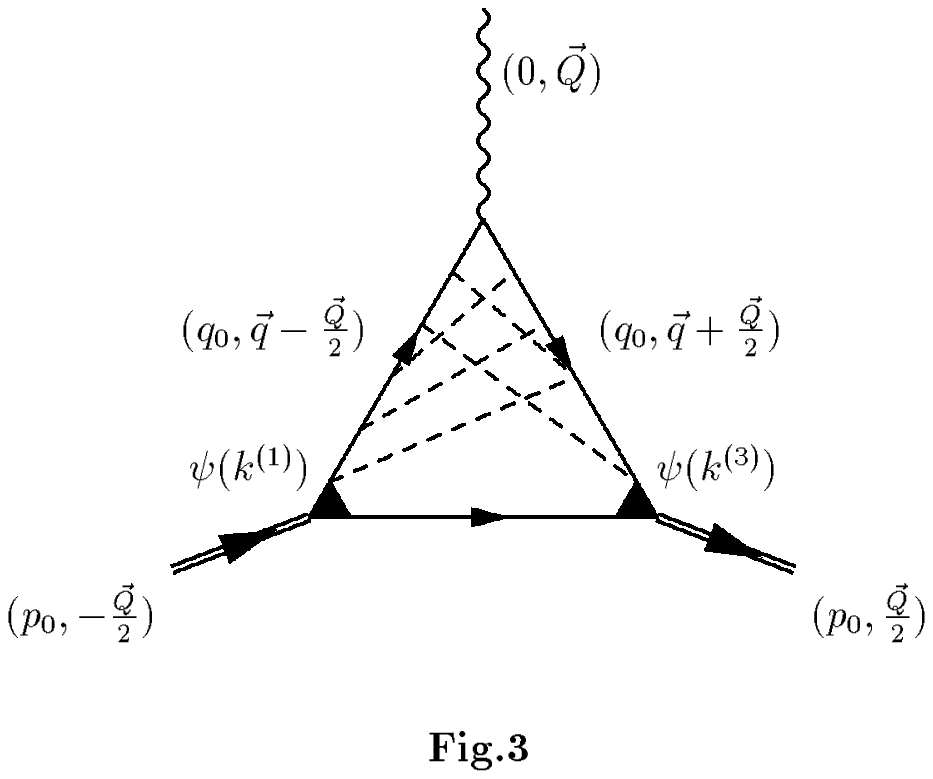}
\end{figure}


\begin{thebibliography}{99}
   \bibitem{1}
Landau Course of Theor. Physics, v.IV\\
V.B.Berestetsky, E.M.Lifshitz, L.P.Pitaevski, Quantum
Electrodynamics, Moscow, Nauka, 1980
\bibitem{2} Yu.L.Dokshitzer, D.I.Dyakonov, S.Z.Troyan, Phys. Rep.
{\bf 58} (1980) 271\\
A.H.Mueller, Phys. Rep. {\bf 73} (1981) 237.
\bibitem{3}
V.V.Sudakov, ZhETF {\bf 30}(1956) 87
\bibitem{4} V.G.Gorshkov, Uspekhi {\bf 110} (1973) 45
\bibitem{5}
J.M.Cornwall, G.Tiktopoulos Phys. Rep. {\bf D13} (1978)
3370\\
J.Frenkel, J.C.Taylor, Nucl. Phys. {\bf B116} (1976)
185\\
E.C.Poggio, G.Pollace, Phys. Lett {\bf B71} (77) 135\\
S.D.Ellis and W.J.Stirling, Phys. Rep. {\bf D23} (1981)
214
\bibitem{6}
Yu.A.Simonov in: Perturbative and Nonperturbative
Aspects of Quantum Field Theory, H.Latal, W.Schweiger
(Eds), Lecture Notes in Physics, v. 479, Springer, 1996.
\bibitem{7}
V.A.Fock, Izvestiya Akad. Nauk USSR, OMEN, p. 557 (1937)
\bibitem{8}
R.P.Feynman, Phys. Rev. {\bf 80} (1950) 440\\
J.Schwinger, Phys. Rev. {\bf 82} (1951) 664
\bibitem{9}
Yu.A.Simonov, Nucl. Phys. {\bf B307} (1988) 512, ibid
{\bf B324} (1989) 67\\
Yu.A.Simonov and J.A.Tjon, Ann. Phys. {\bf 228 } (1993)
1
\bibitem{10} G.A.Milekhin and E.S.Fradkin, JhETF {\bf 45} (1963)
1926\\
E.S.Fradkin, Trudy FIAN, {\bf 29} (1965) 7
\bibitem{11} A.I.Karanikas, C.N.Ktorides, Phys. Lett. {\bf B275}
(1992) 403, Phys. Rev. {\bf D52} (1995) 5883.
 \bibitem{12} N.G.van
Kampen, Phys.  Rep.  24 C (1976) 171
 \bibitem{13} V.S.Dotsenko and
S.N.Vergeles, Nucl.  Phys.  {\bf 169} (1980) 527
\bibitem{14}
 A.I.Karanikas, C.N.Ktorides and N.G.Stefanis, Phys. Rev. {\bf
D52} (1995) 5898

\end{thebibliography}
   \end{document}